\documentclass{cjpsupplmod}
\usepackage{epsfig}
\usepackage{amssymb}
\begin{document}
\title{The First Year at LHC: Diffractive Physics}
\authori{M.\,Deile}
\addressi{CERN, Physics Department}
\authorii{on behalf of the TOTEM collaboration\footnote{
V. Avati, V. Berardi, V. Boccone, M. Bozzo, A. Buzzo, 
M.G. Catanesi, S. Cuneo, C. Da Vi\`{a}, M. Deile, K. Eggert, 
F. Ferro, E. Goussev, J.P. Guillaud, J. Hasi, F. Haug, 
R. Herzog, M. J\"{a}rvinen, P. Jarron, J. Kalliopuska, 
K. Kurvinen, A. Kok, V. Kundr\'{a}t, R. Lauhakangas, 
M. Lokaj\'{\i}\v{c}ek, D. Macina, M. Macr\'{\i}, S. Minutoli, 
A. Morelli, P. Musico, M. Negri, H. Niewiadomski, E. Noschis, 
F. Oljemark, R. Orava, M. Oriunno, K. \"{O}sterberg, V.G. Palmieri, 
A.-L. Perrot, E. Radicioni, R. Rudischer, G. Ruggiero, H. Saarikko,
A. Santroni, 
G. Sanguinetti, G. Sette, W. Snoeys, A. Sobol, S. Tapprogge, A. Toppinen, 
A. Verdier, S. Watts and E. Wobst.}}    
\addressii{}
\authoriii{}   \addressiii{}
\authoriv{}    \addressiv{}
\authorv{}     \addressv{}
\authorvi{}    \addressvi{}
\headtitle{The First Year at LHC: Diffractive Physics}
\headauthor{M.\,Deile}
\lastevenhead{M.\,Deile: The First Year at LHC: Diffractive Physics}
\daterec{}
\setcounter{page}{1}
\firstpage{1}
\lastpage{000}
\maketitle

\begin{abstract}
At the LHC, diffractive physics will be explored by the dedicated experiment
TOTEM whose Technical Design Report has been approved in Summer 2004. 
The experimental programme will be carried out partly in TOTEM standalone mode
for purely forward phenomena like elastic scattering,
and partly in collaboration with CMS for processes requiring a full 
rapidity coverage.
ATLAS and ALICE are interested in diffraction for a later stage.

This article presents the TOTEM/CMS running scenario for diffractive physics
in the first year of LHC. We discuss which processes are within reach
and with which statistics they can be measured.
\end{abstract}

\section{Introduction: Acceptance of TOTEM + CMS}
\begin{figure}[h!]
\begin{center}
\vspace*{-5mm}
\epsfig{file=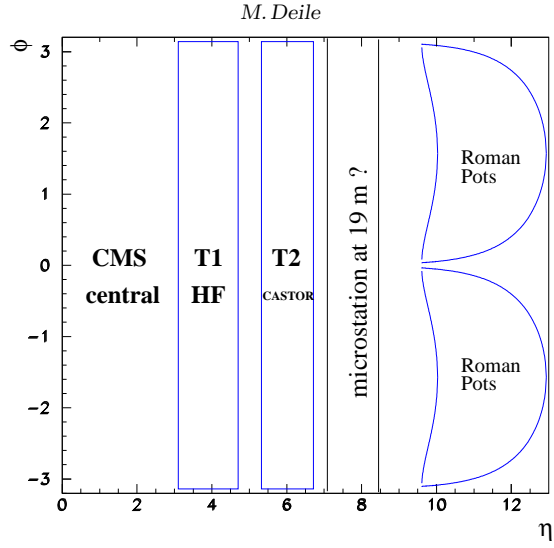,height=7cm}
\caption[*]{Pseudorapidity-azimuth acceptance of the combined TOTEM and CMS
experiments. The microstation at 19\,m is not part of the present design.
}
\label{fig_rapiditycoverage}
\end{center}
\end{figure}

At the interaction point 5 of the LHC the combined CMS and TOTEM detectors 
constitute a
powerful instrumentation for studying diffractive physics.
The TOTEM part contributes the forward trackers T1 and T2 inside CMS
and a system of Roman Pot stations at distances of 147\,m, 180\,m and 220\,m
from the interaction point. 
The TOTEM detectors and their position w.r.t. CMS 
are described in~\cite{fabrizio} (see in particular Fig.~2 and~3 therein), 
and in more detail in the TOTEM TDR~\cite{tdr}. 
The two experiments will be able to take data
together, with TOTEM acting technically as a subdetector of CMS with
the capability to contribute to the level-1 trigger.
The combined CMS+TOTEM experiment has a unique rapidity coverage 
together with an excellent acceptance for leading protons
(Fig.~\ref{fig_rapiditycoverage}). A part of the only coverage gap around
$\eta = 8$ could be filled with an additional leading proton detector
(e.g. a microstation) at a later time.

\begin{figure}[h!]
\begin{center}
\epsfig{file=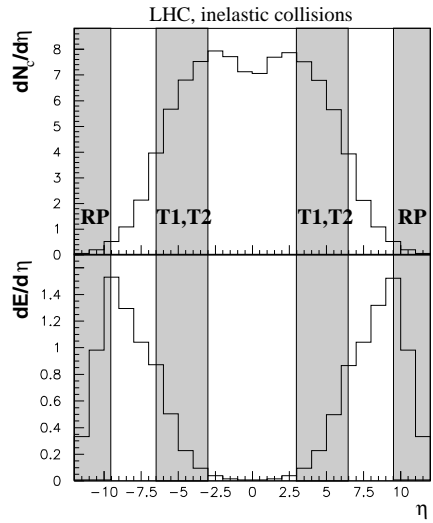,height=7cm}
\caption[*]{Pseudorapidity distribution of the charged particle multiplicity
(upper part) and of the energy flow (lower part) per generic inelastic event. 
}
\label{fig_rapiditydistrib}
\end{center}
\end{figure}
One ingredient for the determination of total cross-section and luminosity 
based on the optical theorem (see Section~\ref{sec_phys_sigmatotlumi}) 
is the measurement of the total elastic and inelastic 
rate.
Minimum bias inelastic events produce particles 
distributed as shown in Fig.~\ref{fig_rapiditydistrib} (upper part). 
The coverage of T1, T2 overlaps
sufficiently with the particle distribution, ensuring an efficiency 
of 99.9\,\% for detecting minimum bias inelastic events.
Elastic events on the other hand are well covered by the Roman Pots. 

Considering that the energy flow in inelastic collisions is predominantly
oriented in the forward direction (Fig.~\ref{fig_rapiditydistrib}, lower part),
the subdetectors T1, T2 and the HF and CASTOR calorimeters are valuable for 
studying forward showers as observed in cosmic ray physics.

\section{The Running Scenarios}

\begin{table}[ht]  
\begin{center}  
\scalebox{0.99}{
\begin{tabular}{|l|c|c|c|c|}\hline 
Running scenario & 1               & 2      & 3       & 4              \\
\hline
Physics           & low $|t|$ el., & diffraction, 
                                            & large $|t|$
                                                      & hard diffr.    \\
                  & $\sigma_{tot}$, 
                    $\mathcal{L}$, & $\mathcal{L}$ 
                                            & elastic &                \\
                  & soft diffr.    &        &         &                \\
\hline\hline
$\beta^*$ [m]     & 1540           & 1540   & 18      & 172            \\ 
\hline
Number of bunches & 43             & 156    & 2808    & $936 \div 2808$\\ 
\hline
Protons per bunch & $0.3 \cdot 10^{11}$ 
                                   & $(0.6 \div 1.15)$ 
                                            & $1.15 \cdot 10^{11}$ 
                                                      & $1.15 \cdot 10^{11}$ \\
                  &                & $\cdot 10^{11}$
                                            &         &                 \\
\hline
Transverse norm.  & 1              & $1 \div 3.75$ 
                                            & 3.75    & 3.75            \\  
emittance [$\mu$m rad] & & & & \\ 
\hline
beam size at IP [$\mu$m] & 454     & $454 \div 880$
                                            & 95      & 294             \\ 
\hline
beam divergence   & 0.29           & $0.29\div 0.57$ 
                                            & 5.28    & 1.7             \\ 
at IP [$\mu$rad]  &                &        &         &                 \\
\hline
$\frac{1}{2}$ crossing angle [$\mu$rad] 
                  & 0              & 0      & 150     & 150             \\
\hline
$t_{min}$ [GeV$^{2}$] @ 220\,m
                  & $2\cdot 10^{-3}$
                                   & $2\cdot 10^{-3}$
                                            & 1.3     & $2\cdot 10^{-2}$ \\
\hline
$t_{max}$ [GeV$^{2}$] @ 220\,m
                  & 0.6            & 0.6    & 7       & 0.6              \\
\hline
$\mathcal{L}$ [cm$^{-2}$s$^{-1}$] 
                  & $1.6 \cdot 10^{28}$ 
                                   & $2.4 \cdot 10^{29}$ 
                                            & $3.6 \cdot 10^{32}$ 
                                                 & $(1\div 4)\cdot 10^{31}$ \\
\hline
\end{tabular}
}
\caption{TOTEM running scenarios.}
\label{tab_scenarios}  
\end{center}  
\end{table}
The running scenarios for forward physics are listed in 
Table~\ref{tab_scenarios}.
For the precise measurement of proton scattering angles 
on the few $\mu$rad level
a special beam optics scheme with $\beta^{*} = 1540\,$m was developed.
It is characterised by a small beam divergence in the interaction point 
(0.29\,$\mu$rad) and a focal point in the Roman Pot station at 220\,m for
both the horizontal ($x$) and the vertical ($y$) track projections.
In order to avoid parasitical bunch crossings downstream of the nominal
interaction point due to the parallelism and large width ($\sim 0.4$\,mm) 
of the two beams, the number of 
bunches will be reduced from 2808 to initially 43 and later 156. 

Scenario 1 will serve for measuring the elastic
cross-section at low $|t|$ (Section~\ref{sec_phys_elastic}), 
the total cross-section, the absolute luminosity 
(Section~\ref{sec_phys_sigmatotlumi})
and soft diffraction (Section~\ref{sec_phys_diffrac}). 
To minimize the emittance, these runs
will be done with a smaller bunch population.
Scenario 3 was introduced for measuring elastic scattering at large $|t|$.
It offers a $t$-acceptance complementary to Scenario 1 and a higher 
luminosity.

Higher luminosities for (semi-) hard diffraction (scenarios 2 and 4) 
can be reached by increasing the number of bunches and the bunch
population. Some details of the optics for Scenario 4 are still under 
development. 

TOTEM operation with normal LHC beam optics ($\beta^{*} = 0.5\,$m) and 
$\mathcal{L} \sim 10^{33}$\,cm$^{-2}$\,s$^{-1}$ is under study
but not yet part of the official programme. A common CMS+TOTEM 
letter of intent for diffractive physics is in preparation.

\section{The Physics Programme}
\label{sec_phys}

\subsection{Total Cross-Section and Luminosity}
\label{sec_phys_sigmatotlumi}
Extrapolations of the total pp cross-section from existing measurements
at $\sqrt{s} \le 1.8\,$TeV to the LHC CM energy of 14\,TeV 
suffer from the conflicting TEVATRON measurements and 
cover a wide range, typically from 90 to 130\,mb (see Fig.~1 
in~\cite{fabrizio}). Cosmic ray experiments have provided direct measurements 
up to a few tens of TeV but with uncertainties on the 20\,\% level. 
The TOTEM collaboration envisages a precision of about 1\,\% using the 
Optical Theorem:
\begin{equation}
\sigma_{tot} = \frac{16 \pi}{1 + \rho^{2}} \cdot
\frac{dN_{el}/dt |_{t=0}}{N_{el} + N_{inel}}\:,
\end{equation}
The extrapolation of the elastic cross-section to $t=0$ suffers mainly from
insufficient knowledge of the functional form 
(see Section~\ref{sec_phys_elastic}) and from uncertainties in beam energy,
detector alignment and crossing angle. The expected systematic error 
is about 0.5\,\% whereas the statistical error is less than 0.1\,\% for only
10 hours of running. The uncertainty of the total rate ($N_{el} + N_{inel}$) is
given by trigger losses and beam-gas background~\cite{fabrizio}; 
it amounts to 0.8\,\%.
Taking also into account the uncertainty in $\rho = 0.12 \pm 0.02$ which
enters only as a quadratic correction term, 
$\sigma_{tot}$ can be measured with about 1\,\% precision.

The same reasoning applies to the absolute luminosity measurement which is 
based on the same observables:
\begin{equation}
\mathcal{L} = \frac{1 + \rho^{2}}{16 \pi} \cdot 
\frac{(N_{el} + N_{inel})^{2}}{dN_{el}/dt |_{t=0}} 
\end{equation}
Also here, a precision of the order 1\,\% is expected.
$\mathcal{L}$ will be directly measured at TOTEM's running scenarios
1 and 2, i.e. at the relatively low luminosities of 
$1.6 \cdot 10^{28}$cm$^{-2}$s$^{-1}$ and $2.4 \cdot 10^{29}$cm$^{-2}$s$^{-1}$.
In these two operating points the CMS luminosity monitors will be calibrated
against TOTEM's absolute result, in order to be used later for relative 
measurements at higher luminosities.

A different approach is envisaged by the ATLAS collaboration~\cite{atlaslumi}.
It will be attempted to measure the differential cross-section of elastic 
scattering down to $-t < 6 \times 10^{-4}$\,GeV$^{2}$ where the dominating
Coulomb scattering provides the absolute scale. In practice,
the measured event rate would be fitted to
\begin{equation}
\frac{dN}{dt} = \mathcal{L} \pi |f_{C} + f_{N}|^{2} \approx
\mathcal{L} \pi \left|-\frac{2 \alpha}{|t|} + \frac{\sigma_{tot}}{4 \pi} |i + \rho|
{\rm e}^{-b |t| / 2}\right|^{2}
\end{equation}
with $\mathcal{L}$, $\sigma_{tot}$, $\rho$ and $b$ as free parameters.

\subsection{Elastic Scattering}
\label{sec_phys_elastic}
\begin{figure}[!h]
\begin{center}
\epsfig{file=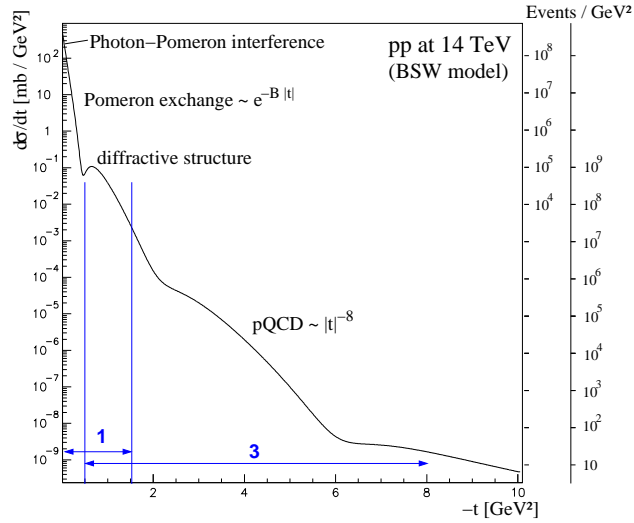,height=7cm}
\caption{
Elastic cross-section predicted by the BSW model~\cite{bsw}. $t$ is the
four-momentum transfer, related to the scattering angle by 
$-t \approx p^{2} \theta^{2}$.
The number of events on the right-hand scales correspond to integrated
luminosities of $10^{33}$ and $10^{37}\,$cm$^{-2}$ (about 1 day with Scenarios
1 and 3).
}
\label{fig_sigmaelastic}
\end{center}
\end{figure}

The differential cross-section $d\sigma/dt$ of elastic scattering according 
to the BSW model is shown in Fig.~\ref{fig_sigmaelastic}. 
Note that there is a rather wide range of predictions for $d\sigma/dt$ at 
LHC-typical centre-of-mass energies.

TOTEM will cover the $|t|$-range from 
$2 \times 10^{-3}\,$GeV$^{2}$ to 8\,GeV$^{2}$ with its running 
scenarios 1 and 3 (Fig.~\ref{fig_sigmaelastic}).
Since the $t$-acceptance for scenario 1 with its $\beta^{*} = 1540$\,m optics 
ends around 1.5\,GeV$^{2}$ (see also Fig.~10 in~\cite{fabrizio}) 
and since at higher $|t|$ more luminosity 
is needed, scenario 3 with a $\beta^{*} = 18$\,m optics was created to extend 
the reach in $|t|$. The ranges accessible to the two scenarios overlap between
0.5 and 1.5\,GeV$^{2}$.
Comfortable statistics can be collected in only 1 day of running with each of
the scenarios (see right-hand scales in Fig.~\ref{fig_sigmaelastic}).
At $|t|<0.1\,$GeV$^{2}$ the statistics are sufficient for narrow bins of 
$10^{-3}$\,GeV$^{2}$ width: at $2 \times 10^{-3}\,$GeV$^{2}$ one will acquire 
$5\times 10^{5}$ events per bin, and at $0.1\,$GeV$^{2}$ still 
$8\times 10^{4}$ events per bin.

Assuming an elastic cross-section of 30\,mb, about $5\times 10^{7}$ events
can be collected during one day at a luminosity of 
$1.6\times 10^{28}$\,cm$^{-2}$s$^{-1}$.

\begin{figure}[!h]
\begin{center}
\epsfig{file=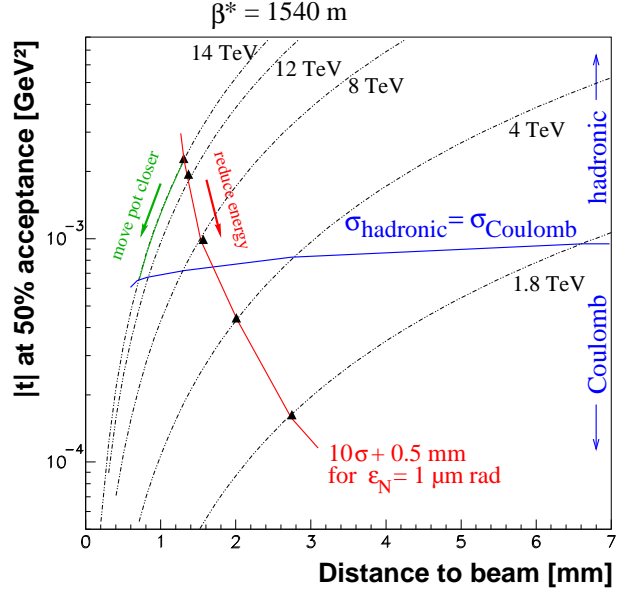,height=8cm}
\caption{$t$-value where 50\,\% acceptance is reached, as a function of the 
detector distance from the beam centre, for different centre-of-mass 
energies. The nominal TOTEM operating point is at 14\,TeV and a normalised 
emittance of 1\,$\mu$m\,rad implying 
a detector-beam distance of 10\,$\sigma + 0.5$\,mm = 1.3\,mm. 
To access the Coulomb region, either the distance between
detector and beam or the energy has to be reduced.
}
\label{fig_tae}
\end{center}
\end{figure}
Access to the Coulomb and interference region 
at $|t| < 2 \times 10^{-3}\,$GeV$^{2}$ will be attempted in two ways
or a combination of them (see Fig.~\ref{fig_tae}):
\begin{itemize}
\item Move the Roman Pot detectors closer to the beam.
This could be achieved
by reducing the number of protons per bunch which would allow for a smaller
emittance.
The luminosity loss could be compensated by increasing the number 
of bunches.
Suppose, the normalised emittance can be reduced to 
$\varepsilon_{N} = 0.5\,\mu$m\,rad, and
stable LHC operation allows the detectors to approach the beam to 
$10\,\sigma + \delta$ with $\delta = 0.1\,$mm. 
Then the $|t|$-value with 50\,\% acceptance would 
be given by
\begin{equation}
\label{eqn_t50}
|t_{50}| = \frac{2\,p^{2}}{L_{\rm eff,\,y}^{2}} \left(10 \sqrt{\frac{\varepsilon_{N} \beta_{y}}
{\gamma}} + \delta\right)^{2} = 5.9\times 10^{-4}\,{\rm GeV}^{2} \:,
\end{equation}
where $p = 7\,$TeV is the proton momentum, $L_{\rm eff,\,y} = 272\,$m is the 
beam optics' effective length in the vertical plane, $\beta_{y} = 48\,$m is 
the betatron function at the detector, and the relativistic $\gamma$ is 7460.5.
\item 
Run the LHC with a reduced centre-of-mass energy $\sqrt{s} = 2\,p \le 6\,$TeV.
According to (\ref{eqn_t50}), the reduction of $|t_{50}|$ with decreasing $p$
due to the leading factor $p^{2}$ is lessened by the 
$\frac{1}{\sqrt{\gamma}} \propto \frac{1}{\sqrt{p}}$ dependence of the beam 
width $\sigma$. The resulting effect behaves like
$|t_{50}| \propto p^{2} \left(\frac{C^{2}}{p}+2 \frac{C \gamma}{\sqrt{p}}+ \delta^{2}\right) = C^{2} p + 2\,C\,\gamma\,p^{3/2} + \delta^{2} p^{2}$.
\end{itemize}

\subsection{Diffraction}
\label{sec_phys_diffrac}
\subsubsection{Diffraction at $\beta^{*} = 1540\,$m}
\begin{figure}[h!]
\begin{center}
\epsfig{file=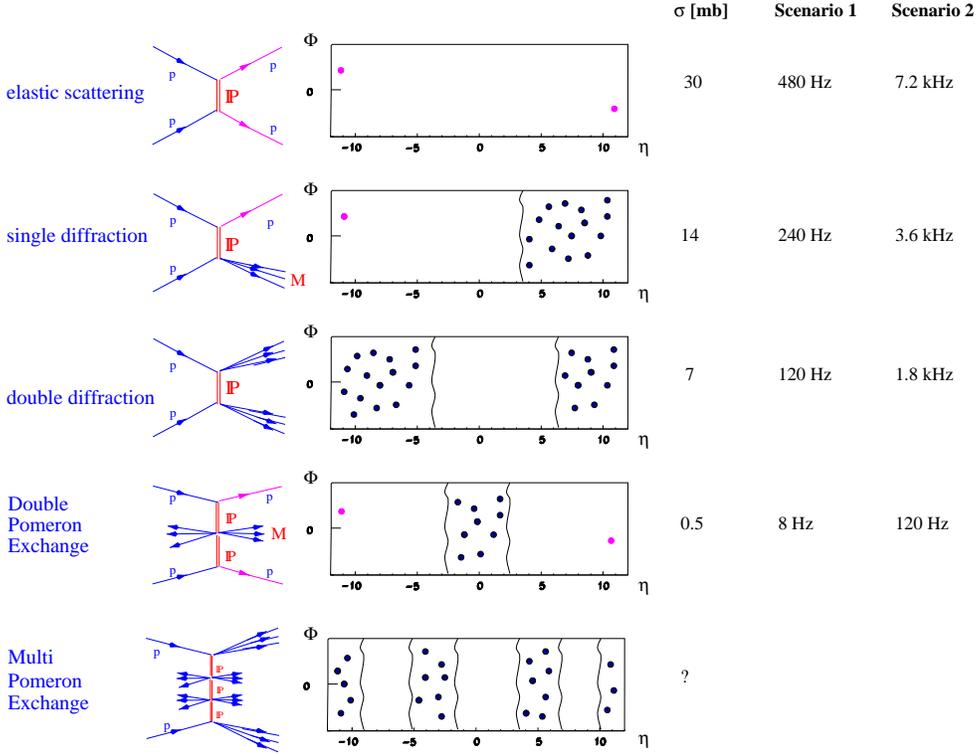,height=10cm}
\caption{Diffractive process classes (left) with their typical particle 
distributions
in the pseudorapidity--azimuth plane (middle)
and their expected event rates with the two
high-$\beta^{*}$ running scenarios (right). Scenario 1 has 
$\mathcal{L} = 1.6 \cdot 10^{28}$\,cm$^{-2}$s$^{-1}$ and scenario 2 has
$\mathcal{L} = 2.4 \cdot 10^{29}$\,cm$^{-2}$s$^{-1}$.
}
\label{fig_diffracprocesses}
\end{center}
\end{figure}

The data taken during 1 day with scenario 1 contain not only 
$4.8\times 10^{7}$ elastic events and $10^{8}$ non-diffractive inelastic 
events, but
also about $3.4\times 10^{7}$ (mostly soft) diffractive events.
Fig.~\ref{fig_diffracprocesses} shows the basic classes of diffractive 
processes and the expected 
statistics with scenarios 1 and 2. The latter has the advantage
of providing a 15 times higher luminosity which can serve for processes
with small cross-section, such as Double Pomeron exchange or semi-hard
phenomena.

Note that the estimates of diffractive cross-sections are 
based on rather uncertain extrapolations from lower energies to LHC
conditions. 
The simultaneous observation of rapidity gaps $\Delta \eta$ and the 
momentum loss $\xi \equiv \Delta p/p$ of surviving protons will allow to
test the relationship $\Delta \eta = -\ln \xi$, valid for pure 
colour singlet exchange. It is still unknown
with which probability the rapidity gaps will ``survive'' at LHC energies, 
i.e. not be filled by hadronisation from 
higher-order soft rescattering processes.

\begin{figure}[h!]
\begin{center}
\epsfig{file=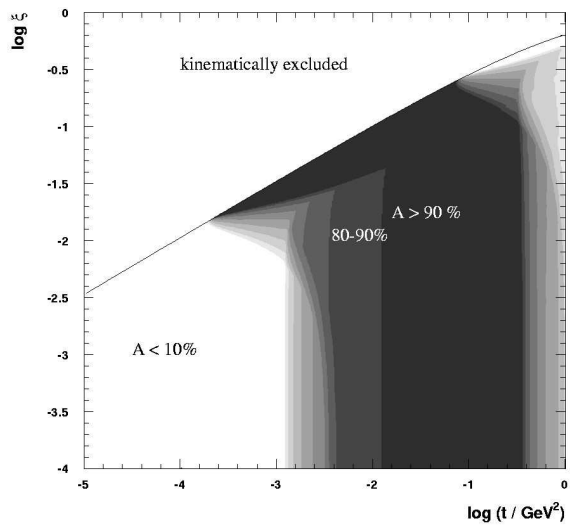,height=7cm}
\caption{
Acceptance for diffractive protons with the 1540\,m optics 
in the RP station at 220\,m. 
}
\label{fig_txiaccept}
\end{center}
\end{figure}

The great advantage of the scenarios at $\beta^{*} = 1540\,$m is the 
very good acceptance in momentum transfer $t$ and momentum loss 
$\xi \equiv \Delta p/p$ (Fig.~\ref{fig_txiaccept}). For 
$|t|>0.002$ all protons are detected independent of $\xi$. 
Assuming a differential cross-section 
\begin{equation}
\frac{d\sigma}{d\xi\, dt} \propto \frac{1}{\xi} {\rm e}^{-B |t|}
\end{equation}
with $B = 5.6\,$GeV$^{-2}$, the total acceptance (integrated over
$t$ and $\xi$) is about 95\,\%.

The $\xi$ 
resolution using the RP stations at 147\,m and 220\,m with the dipole D2 in
between is about $5\times10^{-3}$. 

\subsubsection{Diffraction at Intermediate $\beta^{*}$}
For double-pomeron exchange and hard diffractive processes with their lower 
cross-sections, scenario 4 with a luminosity of the order
$10^{31}$\,cm$^{-2}$s$^{-1}$ will be advantageous. Although some details 
of that scenario are still under study, the acceptance in $t$ and $\xi$ 
is expected to be about 86\,\%.
For the $\xi$-resolution a level of $\lesssim 10^{-3}$ will be 
attempted, i.e.
an improvement of a factor 5 w.r.t. the $\beta^{*}=1540\,$m optics.\\

\begin{figure}[h!]
\begin{center}
\epsfig{file=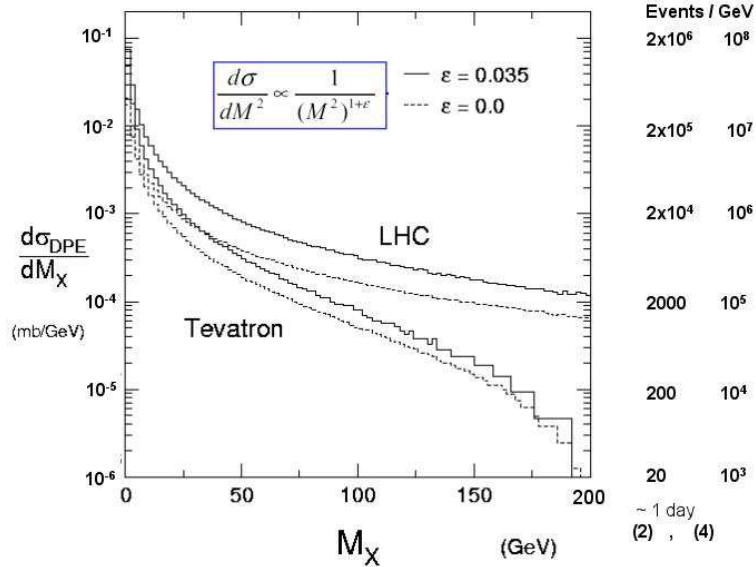,height=8cm}
\vspace*{-5mm}
\caption{Differential cross-section of double-pomeron processes and the 
expected number of events per day for scenarios 2 
($\mathcal{L} = 2.4 \cdot 10^{29}$\,cm$^{-2}$s$^{-1}$) and 
4 ($\mathcal{L} \sim 10^{31}$\,cm$^{-2}$s$^{-1}$).
The continuous and dashed curves represent two models with different 
Pomeron trajectories $\alpha(t) = 1 + \varepsilon + \alpha' t$.
}
\label{fig_dpecross}
\end{center}
\end{figure}

\paragraph{Double Pomeron Exchange (DPE)\\}
Figure~\ref{fig_dpecross} shows the differential cross-section 
of double-pomeron processes extra\-polated to LHC energies.
The event numbers per day (right-hand scales) show by extrapolation 
that diffractive systems 
with masses up to about 2\,TeV are well within reach.

A particularly interesting class of double-pomeron events is exclusive 
central production. It is characterised by a clean signature of two 
surviving protons, two rapidity gaps and only a single particle (or a dijet)
in the diffractive system. Due to the exchange of colour-singlets (having
the quantum numbers of the vacuum), the states produced in the centre must 
obey selection rules on spin $J$, parity $P$ and charge conjugation 
$C$~\cite{kmr_rules}:
\begin{equation}
J^{P} = 0^{+}, 2^{+}, 4^{+}; J_{z} = 0; C = +1
\end{equation}
(in the limit of $t = 0$). 
The $J_{z} = 0$ rule strongly suppresses gg$\rightarrow {\rm q}\bar{\rm q}$ 
background because of helicity conservation (this background would totally 
vanish for massless quarks).

Some examples for exclusive production are given in 
Table~\ref{tab_exclusiveDPE}.

\begin{table}[h!]  
\begin{center}
\scalebox{0.9}{  
\begin{tabular}{|c|c|c|c|c|c|c|}\hline 
Diffractive         & $\sigma$ & Decay channel & 
      BR            & \multicolumn{3}{c|}{Rate at $\mathcal{L}\,[{\rm cm}^{-2}{\rm s}^{-1}] =$}\\
system         &          &               &    
                    & $2.4 \times 10^{29}$ & $10^{31}$ & $10^{33}$ \\
\hline\hline
$\chi_{c0}$    & 3\,$\mu$b & $\gamma J/\psi \rightarrow \gamma \mu^{+} \mu^{-}$&
 $6 \times 10^{-4}$ & 1.6 / h            & 65 / h    & 6480 / h  \\
(3.4\,GeV)     &           & $\pi^{+} \pi^{-} K^{+} K^{-}$ &
 0.018              & 47 / h             & 1944 / h  & 54 / s\\
\hline
$\chi_{b0}$    &  4\,nb   & $\gamma Y \rightarrow \gamma \mu^{+} \mu^{-}$&
 $\le 10^{-3}$          & $\le$ 0.07 / d      & $\le$3 / d     & $\le$ 300 / d \\
(9.9\,GeV)     &          &                                              &
                    &                    &           &         \\
\hline
H (SM)         & 3\,fb    & $b \bar{b}$                                  &
 0.68               & 0.02 / y           & 1 / y     & 100 / y \\
(120\,GeV)     &          &                                              &
                    &                    &           &         \\
\hline
\end{tabular}
}
\caption{Examples of exclusive DPE processes (p + p $\rightarrow$ p + X + p).
The rates do not account for any acceptance or analysis cuts. For 
cross-sections see e.g.~\protect\cite{kmr,kmr_chi}.}
\label{tab_exclusiveDPE}  
\end{center}  
\end{table}
The $\chi_{c}$ resonance may be within reach. The $\chi_{b}$ resonance is
more uncertain because the branching ratios of its decays are not known.
Detecting a 120\,GeV Higgs requires luminosities higher than the ones of
the running scenarios established so far. For this purpose,
new scenarios for standard LHC optics ($\beta^{*} = 0.5\,$m) and 
luminosities ($\mathcal{L} \sim 10^{33}\,{\rm cm}^{-2}{\rm s}^{-1}$) are being
developed for a later stage. These ideas also involve additional 
RP stations in the cryogenic LHC region (at 308\,m, 338\,m and 420\,m from the
IP) in order to achieve an adequate $\xi$ acceptance. \\

\paragraph{Hard Diffraction\\}
If a diffractive event involves a hard subprocess, the diffractive system
(`M' in Fig.~\ref{fig_diffracprocesses}) will contain jets. As an example,
Table~\ref{tab_harddiff} gives the cross-sections and production rates for
dijets in DPE.

\begin{table}[h!]  
\begin{center}
\scalebox{0.9}{  
\begin{tabular}{|c|c|c|c|}\hline 
DPE Dijets & $\sigma$ & \multicolumn{2}{c|}{Rate at $\mathcal{L}\,[{\rm cm}^{-2}{\rm s}^{-1}] =$}\\
($E_{T} > 10$\,GeV) &         & $2.4 \times 10^{29}$ & $10^{31}$ \\
\hline\hline
inclusive    & 1\,$\mu$b   & 864 / h  & 10 / s    \\
\hline
exclusive    & 7\,nb       & 6 / h    & 252 / h   \\
\hline
\end{tabular}
}
\caption{Cross-sections and rates for inclusive and exclusive dijets production
by double-pomeron exchange~\cite{kmr,petrov}.}
\label{tab_harddiff}  
\end{center}  
\end{table}
\section{Conclusions}
A menu for elastic and diffractive physics in the first year of LHC has
been outlined. Starting with several runs of about 1 day with the dedicated
$\beta^{*} = 1540\,$m and 18\,m beam optics, a data sample containing
about $10^{8}$ minimum bias events, $4.8 \times 10^{7}$ elastic events
and $3.4 \times 10^{7}$ diffractive events will be collected.
These data will serve for the measurement of the total cross-section and 
the luminosity with about 1\,\% precision. The elastic 
differential cross-section will be studied in a momentum transfer range from
$10^{-3}\,$GeV$^{2}$ to 8\,GeV$^{2}$. In addition, attempts will be made to 
reach the Coulomb/nuclear interference region at 
$t \lesssim 6 \times 10^{-4}\,$GeV$^{2}$ by running LHC at a lower energy
$\sqrt{s} \lesssim 6\,$TeV or with reduced emittances.

The data already recorded at that stage will also allow studies of
soft diffraction. Expected are $2.4\times 10^{7}$ single diffractive, 
$1.2\times 10^{7}$ double diffractive and $0.1\times 10^{7}$ double Pomeron 
events.

To make semi-hard diffractive processes 
(involving transverse momenta $> 10\,$GeV)
accessible, the number of bunches and the bunch population will be increased
to yield a 15 times higher luminosity, while keeping the optics unchanged.

For hard diffraction and rare DPE phenomena an additional optics configuration 
with $\beta^{*} = 170\,$m and a luminosity of 
$(1 \div 4) \times 10^{31}\,{\rm cm}^{-2}{\rm s}^{-1}$ is under development.

In the farther future there will also be a diffractive programme for standard
LHC running conditions with the $\beta^{*} = 0.5\,$m optics and luminosities
$\sim 10^{33}\,{\rm cm}^{-2}{\rm s}^{-1}$, possibly giving access to very rare
processes like exclusive Higgs production.

\end{document}